\documentclass[journal]{IEEEtran}
\usepackage{graphicx}  % Written by David Carlisle and Sebastian Rahtz
\usepackage{graphicx}
\graphicspath{{../eps/}{../ps/}}
\usepackage{psfrag}    % Written by Craig Barratt, Michael C. Grant,
\usepackage{url}       % Written by Donald Arseneau
\begin{document}
\title{User Profile Based Research Paper Recommendation}


\author{Harshita Sahijwani% <-this % stops a space
	%\thanks{Manuscript received January 20, 2002; revised November 18, 2002.
       % This work was supported by the IEEE.}% <-this % stops a space
	%\thanks{M. Shell is with the Georgia Institute of Technology.}
	\emph{DA-IICT,\ Gandhinagar}
	201301179@daiict.ac.in
	\newline \textbf{Supervisor}
	\emph{Sourish Dasgupta}
	}
\maketitle
\begin{abstract}
Recommender Systems (RSs) are software tools and techniques providing suggestions for items to be of use to a user. [1]
For a recommender system to be effective, it has to have a comprehensive and appropriately catalogued repository of resources. Also, it has to be accurate in understanding the need of the user, and basically has to correctly profile the user. Ideally, one needs to take into consideration the fact that a user's preferences are constantly changing and that is often not done by recommender systems. To meet this need, we propose an interactive research paper recommender system that observes the various themes that occur in a researcher's work over time and keeps learning the user's preferences constantly based on the user's feedback of the papers recommended to them.
We use a topic model to take into account the themes occurring in the document. The query is modeled as a bag of topics, where topics represent latent generative clusters of relative words. The entire collection or research papers is also modeled this way. We then estimate the similarity between the query and each research paper using a bag-of-topics based similarity measure and find the best ones. To take into account the user’s preferences, we keep track of the papers which the user likes and augment the query with the topics which recur in a lot of the user’s preferred papers the next time recommendations are required. We also truncate the topics which the user seems to ignore. 

\end{abstract}

% Will be ignored for other modes.
\IEEEpeerreviewmaketitle

\section{INTRODUCTION}
% \PARstart{A}{demo} file is ....
% \PARstart{A}{}demo file is ....
% \PARstart{T}{his demo} file is ....
	%\PARstart{}{}
This document is a template. An electronic copy can be downloaded from the BTP website (\url{https://uspmes.daiict.ac.in/btpsite}).  For questions on report writing guidelines, please contact your Respective supervisor. Off-Campus students may contact their respective On-Campus mentors.
% May all your publication endeavors be successful.

 
	%\hfill November 18, 2002
\section{Related Work}
 For research papers, many innovative recommender systems have been created. Babel [2] uses a network based approach to classify papers into classic papers which everyone must read and papers meant for experts in the field. For user profiling, approaches like  representing the user as a feature vector which is a function of the references and citations of their most recent paper [3] have been employed. As described in [6], a variety of techniques have been applied to solve the problem of research paper recommendation which involve collaborative filtering, content based filtering and graph based approaches. ActiveCite [7] is an interactive system like our recommender system. It suggests citations to the user while they are working on their paper. But its goal is to assist the user in writing rather than to assist the user in actual \textit{research}. Recommendation on the basis of \textit{long term} observation of a user's preferences in terms of \textit{themes} has not been done by any of these systems.
\section{BACKGROUND}
\subsection{Problem Statement}
Given a query Q describing a user's project description and a set of research papers R, find a subset S of R that has research papers most relevant to the user. The user is expected to select a set of preferred papers P, which serves as an indicator of their preferences, which are taken into account when the user fires a query again.
Relevance of a research paper is measured on the basis of two things:
\begin{itemize}
\item The document similarity of the query Q and the research paper (to be denoted as r) in question.
\item The document similarity of the user's preferred papers P and the query Q.
\end{itemize}

%\jl{The section as it is now is too short and can be merged into preliminaries.}\sdg{In my opinion, problem statement inside preliminary may be bit confusing in terms of readability?}

%As a motivating instance of the problem, let us consider the two short documents:%\\\todo[inline]{Michael Roeder: As far as I see, these two documents are never used as examples. Before their first usage, they are redefined in the subsection 'WordNet based ...'. Thus, they might be removed from here.}
%\textbf{Document 1}: \textit{``The quick brown fox jumped over a lazy dog.
%The boy was clever to have dodged the fox in the chase"}.\\
%\textbf{Document 2}: \textit{``The boy was very clever.
%He came first in the class, and also played sports dodging the ball promptly.
%The dog lay in the side watching the game."} \\

\subsection{Probabilistic Topic Modeling}
\label{ldamodel}
Documents can be represented as BoW, following the assumption of \textit{exchangeability} [4]. The assumption states that if words are modeled as Bernoulli variables, then within any random sample sequence they are conditionally independent, where the word variables are conditioned on a specific set of latent random variables called \textit{topics}. This renders the joint distribution of every sample sequence permutation (i.e.~the document variable) to remain equal, provided the topic  variables are given. In other words, the assumption is that, the order of words representing a document does not matter as long as the topics, which ``\textit{generates}" the occurrence of words, are known. However, interestingly, these topics are hidden (in terms of their distributions) and hence, we need a mechanism to discover (i.e.~learn) them. This learning process is called \textit{topic-modeling}. In this paper, we use a widely adopted probabilistic topic-modeling technique, called \textit{Latent Dirichlet Allocation}, that involves an iterative Bayesian topic assignment process via variational inferencing, over a train-corpus. The number of topics (and other related hyperparameters) needs to be preset. The prior distribution of topics over documents (and also, words over topics) is taken as Dirichlet. The process results in groupings of words that are related to each other \textit{thematically} (in the distributional semantics sense). As an example, ``\textit{house}" and ``\textit{rent}", after the learning process, might be within the same topic.[5]
\subsection{Vector Space Models}
As [7] proposes, documents can be represented as \textit{vectors}. The bases of the vector space can be index terms. The frequencies of those terms act as the weights of the bases in the vector representation of each document. Here, each term is thought of as a dimension in the space. Similarity between two documents represented this way can thus be computed as the cosine similarity between the two vectors. The bases need not be index terms, they can be any set of attributes of a document that can be logically assumed to be independent of each other.
\section{APPROACH}
\begin{figure*}[t]
 	\centering
 	
 	\includegraphics[width=1.0\linewidth]{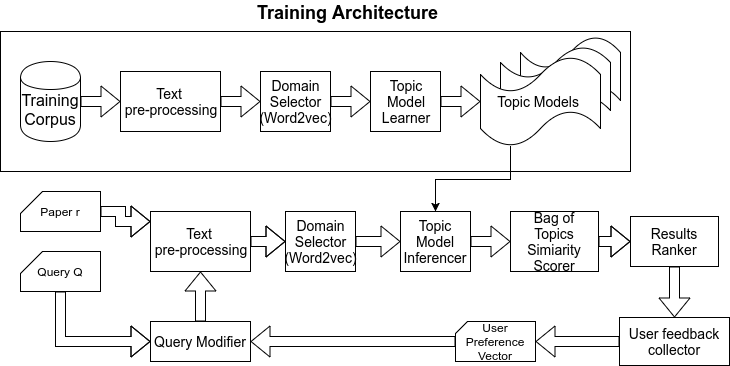}
 	
 	\caption{Recommender System Architecture}
 	\label{architecture} 
 \end{figure*}

The recommender system six core modules:(i) \textit{Domain Selector} (ii) \textit{Topic-Model Learner}, (iii)  \textit{Topic-Model Inferencer}, (iv) \textit{Bag-of-Topics Similarity Scorer},  (v) \textit{Query Modifier}, and (vi) Results Ranker(see Figure~\ref{architecture}).
Instead of relying on just one topic model to represent documents as bags of topics, we train a set of LDA models for documents of different domains. If the training corpus contains documents from the mathematics, humanities and science domains, it is the Domain Selector's job to label the documents with their appropriate domains' tags so that the Topic Model Learner knows which documents need to be used for the training of which model.
The Topic-Model Learner module receives a set of training document corpora, and encodes each training document into a $n$-dimensional vector $[p_1 , \ p_2, \ p_3 \ ...p_n]$. Here $n$ represents the number of topics in the trained LDA model and each value $p_i$ represents the probability of the document to have $i^{th}$ topic. [5]
After the trained model is generated, it is then used by the Topic-Model Inferencer to represent a given document (query or paper) as a \textit{bag} of latent topics using the vector representation of documents and a word-to-topic inverted index.
% \todo[inline]{PT: Changed from word-to-topic mapping function to word-to-topic inverted index. Since we don't define a word-to-topic mapping function formally anywhere}
To compute similarity between the query and a paper, their corresponding topic-sequence representations are fed into the Bag of Topics Similarity Scorer, which makes user of a topics-based vector space to compute the similarity. This process is repeated for all the documents and then they are ranked based on their scores and displayed to the user. The user will indicate preference for some of them. The job of the Query Modifier is to take account of these preferences and modify the next query that the user fires based on these preferences. Sentence-level Similarity Scorer, which uses an adaptation of Smith-Waterman alignment algorithm (discussed in section 3.3). For every mismatch during alignment computation, the algorithm uses Token-level Similarity scorer as a novel compensation computing module that helps to evaluate the degree of mismatch between two topics. %It uses cosine similarity between topic-to-vec representation of every topic, where topic-to-vec is the average word-embedding (i.e.~a high-dimensional vector represention ~\cite{pennington2014glove}) of top-k most probable words in that topic. 

 \subsection{Domain Selector}
 We utilized a Word2vec model for identifying the domain of each paper. 
 A word2vec model represents each word as a vector. It trains itself based on a neural networks based approach on a training corpus. This module computes a representative vector for each domain as the sum of the most significant words in the documents known to be of this domain.
 A new document's domain is assigned based on its closeness to these representative domain vectors. A document can belong to more than one domains if its cosine similarity to the representative domain vector is high enough according to a threshold set by us.
\subsection{Topic-Model Learner}
\label{ldamod}
This is a training-phase module that learns topic-distributions from each document (and thereby learns the word-distribution for each topic) in the train-corpus. 
We use Latent Dirichlet Allocation (LDA) (Section~\ref{ldamodel}) based topic modeling for our purpose. 
It is to be noted that an LDA-based topic model is more accurate when trained over a fixed domain that has a particular vocabulary pattern (i.e.~domain-specific linguistic variations and jargon). 
For instance, a topic model trained over documents from the area of computer science cannot be used to accurately generate topic distributions of documents containing travel blogs. However, it might be able to perform relatively better in related fields such as electrical engineering or statistics or mathematics. 
That is why this module generates a separate topic model for each domain indicated by the Domain Selector.
The Topic-Model Learner first performs text pre-processing on the train-corpus which includes tokenization, lemmatization, and stop-word removal. 
This pre-processing ensures that the LDA model is trained over a condensed natural language text devoid of words which add little or no semantic value to any document. 
All the pre-processing tasks are done using Spacy\footnote{https://spacy.io/}. 
The train-corpus documents are then passed through Gensim's\footnote{https://radimrehurek.com/gensim/} implementation of LDA to learn topic-distributions for the documents. 

This module also creates, for each domain-wise topic model, an inverted topic-word distribution index that maps each word of the vocabulary to topics, along with the probability of that word in the corresponding topic. Its utility is explained in the section below.

\subsection{Topic-Model Inferencer}
\label{ldainf}
%cleanup; topic modeling; sequence generation; segmentation
This is an inferencing phase module. When each document of an unseen document pair is fed into the module, it first performs the same NLP pre-processing as the Topic-Model Learner module. After that, it performs voice normalization on every sentence in the documents, thereby converting passive sentences into their active form. Without this normalization step, the thematic flow of similar sentences (and hence, documents) will appear different even if they have the same semantic content.

The domains for this document are identified. Then for each domain, the cleaned document pair is fed into Gensim's trained LDA topic model for that domain to infer topic distributions of the documents. 
Thereafter, the module transforms the documents into their \textit{bag of topics} representations. 
The word-to-topic mapping is done by using the inverted topic-word distribution index (described in the previous section) where, as the document is passed through the model, every word in the document is assigned the maximum probable topic. 
The generated bag of topics represents the \textit{semantic themes} present in the document content.
\section{Bag of Topics Similarity Scorer}
This module computes the document similarity score of the query and a research paper given their bag of topics representations. The similarity measure that we use is Cosine Similarity between the bag-of-topics vectors of the two documents. For example, if the query and the research paper have the following vec representations, 
\begin{equation}
q = 5 MAT1 + 6 CS3 + 2 HUM4
\end{equation}
\begin{equation}
r = 3 MAT1 + 2 CS3 + 2 ENG5
\end{equation}
The cosine similarity would be 
\begin{equation}
    \sigma = q.r/|q||r| = 0.81
\end{equation}

This module assumes all the topics of all the modules to be linearly independent. This assumption is logical because each domain's topic model is trained on a separate corpus with significantly different terms and their co-occurrences. 
\section{Query Modifier}
This module has the task of learning the user's preferences and modifying the content-based query accordingly.
Given a query q, say,
\begin{equation}
q = 5 MAT1 + 6 CS3 + 2 HUM4,
\end{equation}
it should return a query q', where 
\begin{equation}
q' = 5 MAT1 + 6 CS3 + 2 HUM4 + 0.5 MAT9
\end{equation}
If the user has been showing preference for papers with a high relative frequency of the topic MAT9. 
The query modifier aims to be certain of the user's preference before adding a topic to the query.
For that, a user preference vector is maintained by the module. It basically is a vector representation of the user in the topic space and represents his or her affinity for each topic. An example of a user preference vector would be 
\begin{equation}
u = 0.0005 MAT1 + 0.6 CS3 + 2 HUM4 + -0.3 MAT_2.
\end{equation}
This way, each user occupies a point in the topics vector space. After each N number of queries, where N indicates for how long you want to observe a user before hypothesizing his preferences, the user vector is updated. And for each query, the query is updated as follows. 
\begin{equation}
q' = q + u
\end{equation}
The following example illustrates the motivation behind the algorithm used for updating the user preference vector. 
Suppose the results that the user showed preference for after the query q in equation 1 was fired were 
\begin{equation}
r1 = 4MAT1 + 1CS3 + 3 CS6,
\end{equation}
\begin{equation}
r2 = 5 MAT9 + 1CS3 + 1HUM4,
\end{equation}
Here, r1's most prominent theme is MAT1, which is one of the topics present in the initial query as well. So the weight of MAT1 should be increased. The weight of CS3 should also be increased, but by a lower value because it is the least prominent theme in this document. 
r2 contains CS3, which is one of the topics present in the original query. However, again, it is not one of the prominent themes. And MAT9, which wasn't initially present in the query, is one of the prominent themes. So the weight of MAT9 should be increased in the user preference vector, but not as much as MAT1's, because the user has not shown explicit preference for MAT9 in the query.
\\
Hence we update the user query as follows:
For each topic in each of the results r,
If the topic belongs to the query,
\begin{equation}
u = u + \Delta (1 - relative frequency of topic) topic
\end{equation}

\begin{equation}
u = u + \delta (1 - relative frequency of topic) topic
\end{equation}

$\Delta$ will be a higher value as compared to $\delta$.
If a topic added is added to the query in this manner, but is not shown prefernce for later for many iterations, it is awarded a penalty $\beta$, which is higher thatn $\Delta$.
\section{EVALUATION}
We evaluated the system on a dataset having the queries and the preferred research papers of 15 researchers\footnote{https://www.comp.nus.edu.sg/~sugiyama/Dataset1.html} which was also used for evaluation in [3]. 
We trained twelve topic models for twelve different domains, namely computer science, library and information science, arts, humanities, mathematics, science, leisure, health and medicine, engineering, social science, business and others using documents of the respective topics. 
For domain selection, we made use of a word2vec model trained on Wikipedia documents.

We calculated the Jaccard similarity score of the acquired results and the desired results for both the original query(q) and the query modified after 10 iterations of interaction with the user(q'). We found the Jaccard similarity to be higher for the modified query for 11 out of the 15 users.

\end{document}